\documentclass[journal]{IEEEtran}
\usepackage[utf8]{inputenc}

\usepackage{graphicx}%handle the images
\usepackage{siunitx} %Handle the SI units
\DeclareSIUnit{\dBm}{dBm}
\DeclareSIUnit{\dBc}{dBc}
\DeclareSIUnit{\dB}{dB}

\usepackage{cite} %Prettify the references
\usepackage{doi} %Provide the DOI links
\usepackage{amsmath} %allow multiline equations

\title{Ultra-wideband free-space optical phase stabilisation}

\author{Benjamin P. Dix-Matthews,~\IEEEmembership{Member,~IEEE,} David R. Gozzard,~\IEEEmembership{Member,~IEEE,} Skevos F.E. Karpathakis,~\IEEEmembership{Member,~IEEE,} Charles T. Gravestock and Sascha W. Schediwy
\thanks{
This research was supported by the Australian Research Council's Centre of Excellence for Engineered Quantum Systems (EQUS, CE170100009), and the SmartSat Cooperative Research Centre.
D.R.G. is supported by a Forrest Research Foundation Fellowship.
B.P.D-M. is supported by an Australian Government Research Training Program (RTP) Scholarship and an SmartSat Cooperative Research Centre Top-Up Scholarship.
\textit{(Corresponding author: Benjamin P. Dix-Matthews.)}

Benjamin P. Dix-Matthews, David R. Gozzard, Skevos F.E. Karpathakis, Charles T. Gravestock and Sascha W. Schediwy are with the International Centre for Radio Astronomy Research, The University of Western Australia, Crawley 6009, WA, Australia.

Benjamin P. Dix-Matthews, David R. Gozzard and Sascha W. Schediwy are additionally with Australian Research Council Centre of Excellence for Engineered Quantum Systems, The University of Western Australia, Crawley 6009, WA, Australia. (email: benjamin.dix-matthews@research.uwa.edu.au)
}
}

\usepackage{fancyhdr}
\makeatletter
\def\ps@IEEEtitlepagestyle{
\pagestyle{fancy}

\lhead{Accepted for publication in IEEE Communications Letters - Copyright transferred to IEEE.}
\rhead{}
\cfoot{}
}

\begin{document}
\maketitle

\begin{abstract}
Free-space optical (FSO) communications has the potential to revolutionize wireless communications due to its advantages of inherent security, high-directionality, high available bandwidth and small physical footprint.
The effects of atmospheric turbulence currently limit the performance of FSO communications.
In this letter, we demonstrate a system capable of indiscriminately suppressing the atmospheric phase noise encountered by independent optical signals spread over a range of~\SI{7.2}{\tera\hertz} (encompassing the full optical C-Band), by actively phase stabilizing a primary optical signal at \SI{193.1}{\tera\hertz} (\SI{1552}{\nano\meter}).
We show~\SI{\sim30}{\dB} of indiscriminate phase stabilization over the full range, down to average phase noise at~\SI{10}{\hertz} of~\SI{-39.6}{\dBc\per\hertz} when using an acousto-optic modulator (AOM) as a Doppler actuator, and~\SI{-39.9}{\dBc\per\hertz} when using a fiber-stretcher as group-delay actuator to provide the phase-stabilization system's feedback.
We demonstrate that this suppression is limited by the noise of the independent optical signals, and that the expected achievable suppression is more than~\SI{40}{\dB} greater, reaching around~\SI{-90}{\dB\per\hertz} at~\SI{10}{\hertz}. We conclude that~\si{40~Tbps} ground-to-space FSO transmission would be made possible with the combination of our stabilization system and other demonstrated technologies.

\end{abstract}


\section{Introduction}
\IEEEPARstart{F}{ree-space} optical (FSO) communications provide many advantages over conventional radio-frequency wireless technology, such as greater inherent security, high directionality, higher available bandwidth, and small physical footprint.
These advantages make FSO communication valuable to military, commercial, and scientific applications requiring secure high-speed data transfer over line-of-sight free-space channels.

The efficacy of these FSO channels are limited by their vulnerability to the effects of atmospheric turbulence.
The spatial and temporal fluctuations of the atmosphere vary the propagation of transmitted optical beams, leading to intensity fluctuations and increased phase noise on the received signal~\cite{Robert2016,swann2017low,fante1975electromagnetic}.
The intensity fluctuations are caused by beam-wander and scintillation due to first-order and higher-order variations in the propagating beam.
There is ongoing research into suppressing these effects through the use of adaptive optics and fast steering mirrors~\cite{Lim2019,Vedrenne2019,Ansari2018,Bonnefois2019,Chen2018,moision2017}.
The increased phase noise is due to zeroth-order (or piston-mode) variations imprinting phase perturbations onto the propagating beam.
Overcoming this degradation is important for the stable FSO transfer of coherent optical reference signals.

The stable transmission of coherent reference signals over FSO channels is important to many scientific applications~\cite{grotti2018geodesy,riehle2017optical,Lion2017,Takamoto2020}.
This has led to the development of coherent phase stabilization systems, capable of suppressing the atmospheric phase noise over point-to-point~\cite{dix2020point} and folded FSO channels~\cite{Gozzard2018,Dix-Matthews2020Coherent,Kang2019,chen2017sub,Djerroud:s,djerroud2010coherent}.
These coherent phase stabilization systems are based on similar fiber-based stabilization techniques~\cite{ma1994delivering,foreman2007remote,Schediwy2013}, and are focused on suppressing the atmospheric phase noise experienced by a single optical frequency during FSO transmission.

The phase fluctuations experienced by transmitted optical signals during atmospheric propagation are only weakly dependent on wavelength~\cite{Kang2019}.
This means that measurement of the atmospheric phase fluctuations experienced by a primary optical signal can be used to infer and stabilize the phase noise experienced by independent secondary optical signals separated in wavelength.
In 2019, Kang et al.~\cite{Kang2019}, used a frequency comb-based optical source to demonstrate that this technique was able to stabilize several secondary optical signals offset from the primary optical signal by~\SI{\sim4}{\tera\hertz}.
This ability to use a single primary optical signal to indiscriminately stabilize the atmospheric phase noise encountered by independent signals in an ultra-wide spectral range has important significance for FSO communications using wavelength division multiplexing (WDM)~\cite{Poliak2018,MataCalvo2019,Dochhan2019}.

In this letter, we demonstrate indiscriminate phase stabilization of independent secondary optical signals from a tunable laser in 19 channels spanning~\SI{7.2}{\tera\hertz} (\SIrange{190.0}{197.2}{\tera\hertz} or \SIrange{1578}{1520}{\nano \meter}), by phase stabilizing a primary optical signal at the centre of this range (\SI{193.1}{\tera\hertz} or \SI{1552}{\nano\meter}).
We present results using both a fiber-stretcher as group-delay actuator, and an acousto-optic modulator (AOM) as a Doppler actuator, to provide the phase-stabilization system's feedback.
We show that the indiscriminate phase stabilization of the secondary optical signal suppresses the phase noise by greater than~\SI{\sim30}{\dB}, down to an average phase noise at~\SI{10}{\hertz} of \SI{-39.6}{\dBc\per\hertz} when using an AOM, and~\SI{-39.9}{\dBc\per\hertz} when using a stretcher to provide feedback.
We demonstrate that the measured suppression is limited by the poor coherence of the secondary optical source, and find the achievable suppression is likely to be over~\SI{40}{\dB} greater, down to around~\SI{-90}{\dBc\per\hertz} at~\SI{10}{\hertz}.

\section{Transfer System Description}
\begin{figure*}
    \centering
    \includegraphics[width=\linewidth]{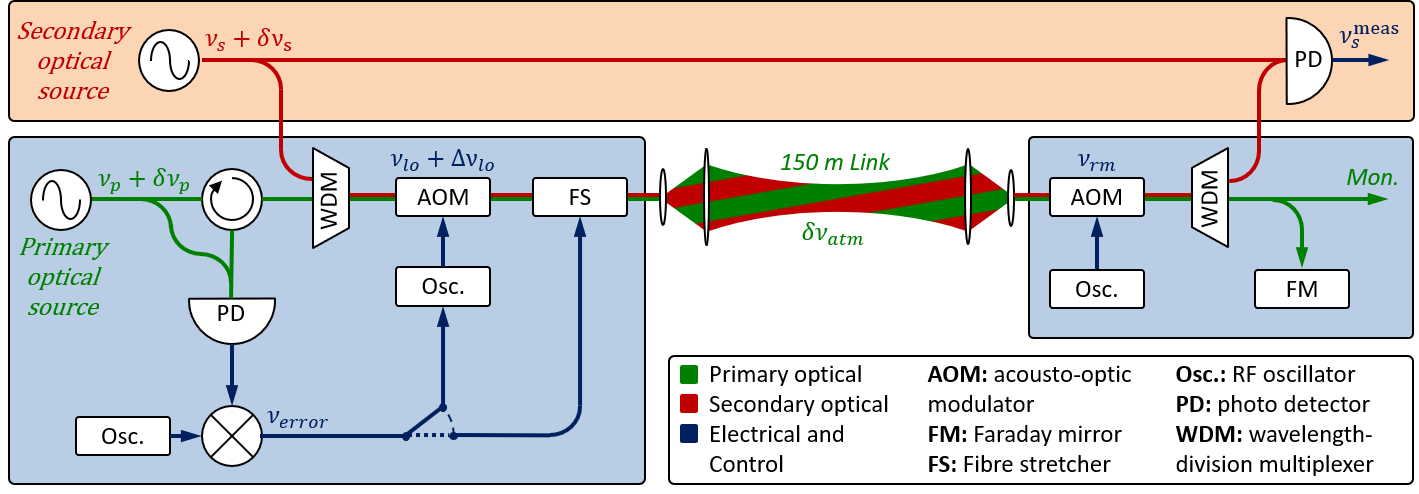}
    \caption{Block diagram of the indiscriminate phase stabilization system. Green lines indicate primary optical signals, Red lines indicate secondary optical signals, and blue lines indicate electrical signals used for control and stability measurements. Noise processes begin with a $\delta$, controlled phase variations begin with a $\Delta$, and both are time dependent.}
    \label{fig:Figure1}
\end{figure*}

The phase stabilization system presented in this paper actively suppresses the atmospheric phase noise encountered by a primary optical signal during transmission over a FSO channel.
A block diagram of the experimental system is shown in Fig.~\ref{fig:Figure1}.
The primary optical signal is produced by an NKT Photonics X15 laser with a nominal frequency, $\nu_p$, of \SI{193.1}{\tera\hertz}, a phase noise of $\delta \nu_{p}(t)$, and an output power of~\SI{15}{\dBm} (\SI{30}{\milli\watt}).
The phase stabilization takes advantage of the weak dependence of the atmospheric phase noise on the optical frequency~\cite{Kang2019} to indiscriminately phase stabilize independent secondary optical signals from a Topica Photonics DL100 Tunable laser with a variable frequency, $\nu_{s}$, and a phase noise of $\delta \nu_{s}(t)$.
The variable frequency of this laser is tuned over~\SI{7.2}{\tera\hertz} (\SIrange{190.0}{197.2}{\tera\hertz}) in order to probe the efficacy of the indiscriminate phase stabilization.

The design of the phase stabilization system is based on an interferometer, with one arm encompassing the FSO channel.
The primary optical signal is passed through a circulator to a wavelength-division multiplexer (WDM), which combines the secondary optical signal from the tunable laser onto the same fiber.
The combined signal is then passed through a local AOM that shifts the frequency of the transmitted signals by a nominal frequency, $\nu_{lo}$, of \SI{75}{\mega\hertz}, which may be tuned by $\Delta \nu_{lo}(t)$.
The combined signal is then sent through a~\SI{40}{\meter} fiber stretcher, fiber to free-space collimator, Gaussian beam expander, and is launched over a~\SI{150}{\meter} turbulent FSO channel with a $1/e^2$ beam radius of~\SI{16.8}{\milli\meter} and divergence of~\SI{29}{\micro\radian}.

The combined signal is received (following a propagation time of $T$) by a separate Gaussian beam expander and free-space to fiber collimator, after having atmospheric phase noise ($\delta_{atm}(t)$) induced by turbulence in the FSO channel.
The combined signal is then passed through a second AOM.
This remote AOM imparts a frequency shift, $\nu_{rm}$, of \SI{-85}{\mega\hertz}, which enables the servo system to distinguish signals reflected from the remote site from other spurious reflections in the link.
A WDM is then used to extract the received secondary optical signal ($\nu_{s}^{rx}(t)$).
\begin{multline}\label{tlOutput}
\nu_{s}^{rx}(t)=\nu_{s} +\delta \nu_{s}(t-T)+\nu_{lo}+\Delta \nu_{lo}(t-T)\\
+\delta \nu_{atm}(t)+\nu_{rm}
\end{multline}

The received primary optical signal is then split, 30\% is used to monitor the primary optical stability.
The remainder is then reflected by a Faraday mirror, back through the remote AOM and over the FSO channel.
This reflected primary optical signal is then received at the transmission site, and passed back through the local AOM and directed by a circulator to a photo-detector where it forms a hetrodyne-beat against the local primary optical signal.

This hetrodyne-beat signal is then mixed with a local oscillator (of frequency $2\nu_{lo}+2\nu_{rm}$) and the down-converted signal is isolated by low-pass filtering.
The resulting error signal ($\nu_{error}$) contains information about the phase noise encountered by the primary optical signal during transmission over the FSO channel ($\delta \nu_{atm}$).
\begin{multline}
\nu_{error}=-\delta \nu_{p}(t)+\delta \nu_{p}(t-2T)+\Delta \nu_{lo}(t)+\Delta \nu_{lo}(t-2T)\\
+\delta \nu_{atm}(t-2T)+\delta \nu_{atm}(t)
\end{multline}

The error signal is driven to zero by a servo loop, which actively alters the frequency of the local AOM ($\Delta \nu_{lo}$), thereby suppressing the additional atmospheric phase noise.
By approximating $\Delta \nu_{lo}(t)+\Delta \nu_{lo}(t-2T)$ as $2\times\Delta \nu_{lo}(t)$, the resulting local AOM frequency shift is:
\begin{multline}\label{trDrive}
   \Delta \nu_{lo}(t)=\frac{1}{2} (\delta \nu_{p}(t)-\delta \nu_{p}(t-2T)\\
   -\delta \nu_{atm}(t-2T)+\delta \nu_{atm}(t)) 
\end{multline}

This is then substituted, with appropriate time delays, into Eq.~\ref{tlOutput} to get the expected secondary optical signal received with indiscriminate phase stabilization engaged.
\begin{multline}
\nu_{s}^{rx}(t)=\nu_{s}+\nu_{lo}+\nu_{rm}\\
 +\delta \nu_{s}(t-T)]+[\frac{1}{2}\delta \nu_{p}(t-T)-\frac{1}{2}\delta \nu_{p}(t-3T)\\
+\delta \nu_{atm}(t)-\frac{1}{2}\delta \nu_{atm}(t-3T)-\frac{1}{2}\delta \nu_{atm}(t-T)
\end{multline}

This received secondary optical signal is beaten against the secondary optical source ($\nu_{s}+\delta\nu_{s}(t)$) onto a measurement photodetector.
The phase noise of this electrical measurement signal ($\nu_{s}^{meas}(t)$) is then used to analyze the stability of the indiscriminate phase stabilization.
\begin{multline}\label{tlMeas}
\nu_{s}^{meas}(t)=\nu_{lo}+\nu_{rm}\\
 +\delta \nu_{s}(t-T)-\delta \nu_{s}(t)+\frac{1}{2}\delta \nu_{p}(t-T)-\frac{1}{2}\delta \nu_{p}(t-3T)\\
+\delta \nu_{atm}(t)-\frac{1}{2}\delta \nu_{atm}(t-3T)-\frac{1}{2}\delta \nu_{atm}(t-T)
\end{multline}

This measurement signal has a nominal frequency of~\SI{10}{\mega\hertz} ($\nu_{lo}+\nu_{rm}$), and noise contributions from the primary optical signal, secondary optical signal, and the atmospheric turbulence.
These three noise contributions are independent, and thus they may each be analyzed separately by applying the identity for combinations of time delayed copies of a noise process (detailed in the appendix of~\cite{Dix-Matthews2020Coherent}).
% \begin{equation}
% z(t)=\sum_i a_i x(t-T_i)
% \end{equation}
% \begin{equation}
% S_z(f)=\left[\sum_i a_i^2 +\sum_i \sum_{i\neq j}a_i a_j cos(2\pi f(T_i-T_j))\right]S_x(f)
% \end{equation}
The expected contribution from each source ($X$) on the measurement signal ($S^{meas}_{X}$) may be expressed in terms of their inherent noise ($S_{X}$):

\begin{equation}
S^{meas}_{s}(f)=\left[2-2cos(2\pi f T)\right]S_{s}(f)
\end{equation}

\begin{equation}
S^{meas}_{p}(f)=\left[\frac{1}{2}-\frac{1}{2}cos(4\pi f T)\right]S_{p}(f)
\end{equation}

\begin{equation}\label{eq:atmScale}
S^{meas}_{atm}(f)=\left[\frac{3}{2}-\frac{1}{2}cos(2\pi f T)-cos(4\pi f T)\right]S_{atm}(f)
\end{equation}

A typical inherent noise model for the primary optical signal ($S_{p}(f)$) was provided by the manufacturer~\cite{NKT}.
The inherent phase noise model for the secondary optical source ($S_{s}(f)$) was obtained by converting frequency noise data provided by the manufacturer~\cite{toptica}.
The noise model for the atmospheric component ($S_{atm}(f)$) is obtained from the unstabilized atmospheric noise, under the assumption that this phase noise will be dominated by the atmospheric effects.

The system produced for this paper has the option of using a fiber-stretcher as group-delay actuator to close the servo loop, instead of driving the local AOM as in Eq.~\ref{trDrive}.
Though the actuation method changes, the effect on the output signal, and the subsequent equations remain the same.

\section{Results}
\begin{figure}
    \centering
    \includegraphics[width=\linewidth]{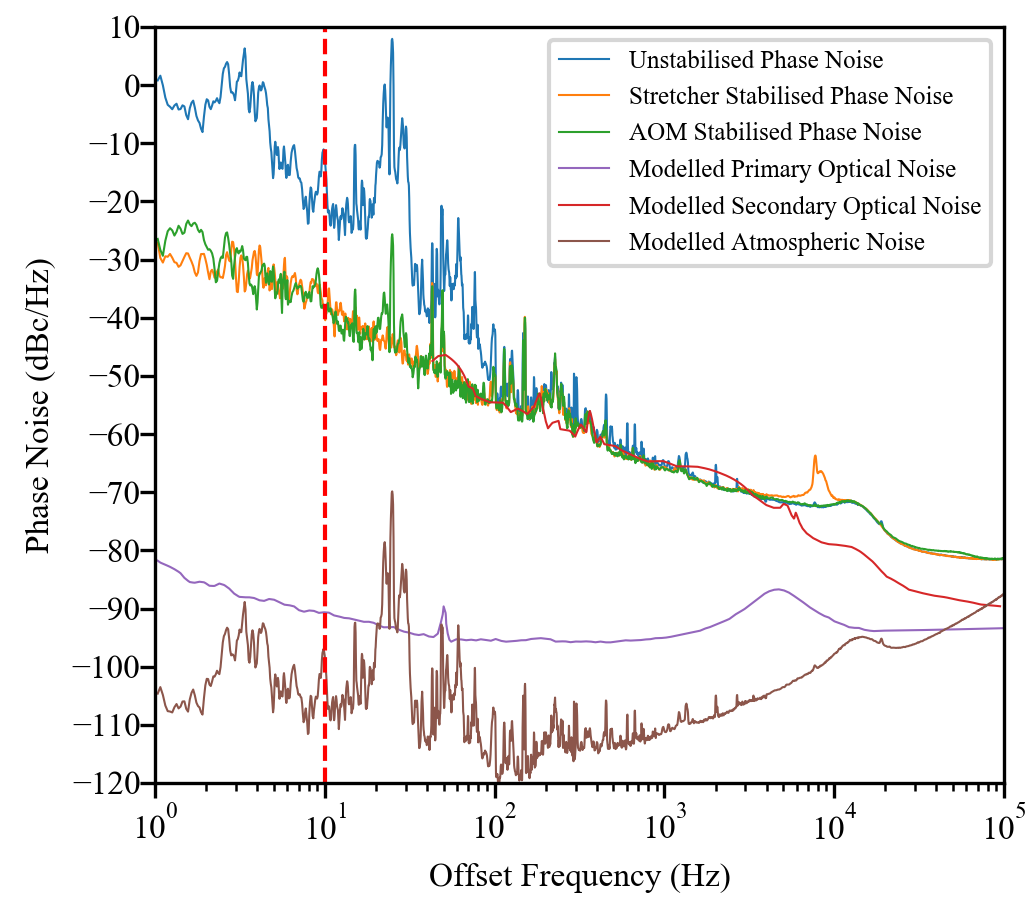}
    \caption{Phase noise of the independent secondary optical signal at a frequency of \SI{197.2}{\tera\hertz} when unstabilized (blue), stretcher stabilized (orange), and AOM stabilized (green). The plot also contains the modeled contributions from the primary optical source noise (purple), secondary optical signal noise (red), and residual atmospheric noise (brown). The vertical dotted line at~\SI{10}{\hertz} indicates the phase noise value corresponding to the results in Fig.~\ref{fig:Figure3}.}
    \label{fig:Figure2}
\end{figure}

The performance of the indiscriminate phase stabilization was determined by measuring the stability of the~\SI{10}{\mega\hertz} measurement signal ($\nu_{s}^{meas}(t)$) using a Microsemi 3120a Phase Noise Test Probe.
Fig.~\ref{fig:Figure2} shows the stability measurements obtained for Channel 72 (\SI{197.2}{\tera\hertz}) when unstabilized (blue), stretcher stabilized (orange), and AOM stabilized (green).
The predicted noise contributions due to primary optical source noise ($S^{meas}_{p}(f)$), secondary optical signal noise ($S^{meas}_{s}(f)$), and residual atmospheric effects ($S^{meas}_{atm}(f)$) are also displayed.

When unstabilized, the stability of the output signal is dominated by atmospheric effects for frequencies below~\SI{100}{\hertz}.
Above~\SI{100}{\hertz} the inherent noise of the secondary optical source dominates.
When stabilized, the atmospheric noise is successfully suppressed, and the inherent phase noise of the secondary optical source appears to dominate even at lower frequencies.

\begin{figure}
    \centering
    \includegraphics[width=\linewidth]{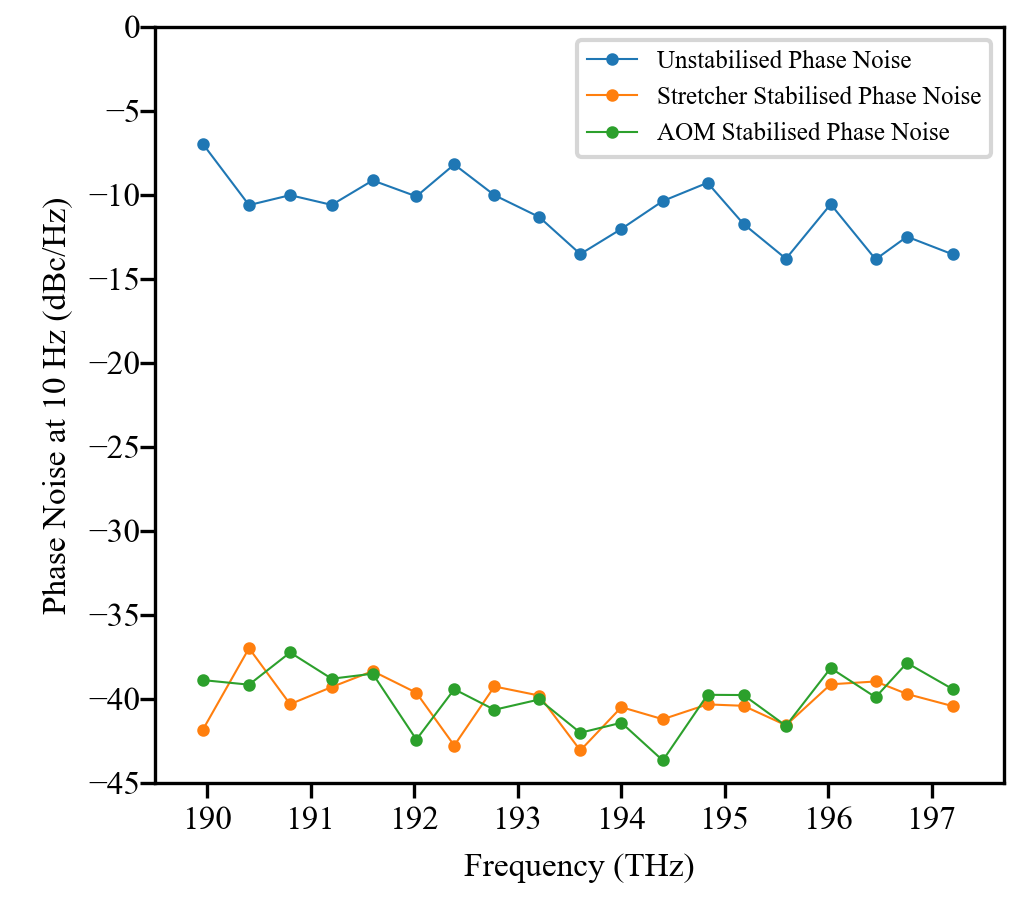}
    \caption{The phase noise at~\SI{10}{\hertz} for each secondary optical frequency measured over the~\SI{7.2}{\tera\hertz} range when unstabilized (blue), stretcher stabilized (orange), and AOM stabilized (green). This indicates the efficacy of the indiscriminate phase stabilization system over a broad spectral range.}
    \label{fig:Figure3}
\end{figure}

Equivalent stability measurements were taken every~\SI{400}{\giga\hertz} across a~\SI{7.2}{\tera\hertz} range centered at \SI{193.1}{\tera\hertz}.
The phase noise at~\SI{10}{\hertz} for the unstabilized (blue), stretcher stabilized (yellow), and AOM stabilized (green) over the~\SI{7.2}{\tera\hertz} range are shown in Fig.~\ref{fig:Figure3}.
At other offset frequencies the phase noise for the unstabilized, stretcher stabilized, and AOM stabilized measurements showed similar consistency over the full~\SI{7.2}{\tera\hertz} range.

While unstabilized, the average phase noise at~\SI{10}{\hertz} is~\SI[parse-numbers=false]{-10.5^{+1.6}_{-2.6}}{\dBc\per\hertz}.
With the phase stabilization engaged, the average phase noise at~\SI{10}{\hertz} decreases to~\SI[parse-numbers=false]{-39.6^{+1.3}_{-1.8}}{\dBc\per\hertz} when using the AOM and~\SI[parse-numbers=false]{-39.9^{+1.3}_{-1.8}}{\dBc\per\hertz} when using the stretcher.
This represents a~\SI{\sim30}{\dB} reduction in phase noise, with insignificant dependence on the type of actuator used.

\section{Discussion}
The results indicate that the phase stabilization system operating at~\SI{193.1}{\tera\hertz} was successfully able to indiscriminately stabilize adjacent channels over a~\SI{7.2}{\tera\hertz} range.
The average phase noise at~\SI{10}{\hertz} was reduced by~\SI{\sim30}{\dB}, down to a value dominated by the inherent phase noise of the secondary optical source.
The modeled noise floors in Fig.~\ref{fig:Figure2} suggest that a reduction in the phase noise of the secondary optical source, to below the primary optical source, would improve the measured suppression by~\SI{>40}{\dB}.
With the phase noise of the primary optical source dominating, the expected phase noise at~\SI{10}{\hertz} would be~\SI{\sim-90}{\dBc\per\hertz}.

The average phase noise after the indiscriminate phase stabilization was equal to within statistical uncertainty, when using either the Doppler (AOM), or group delay (fiber stretcher) feedback actuator.
Some advantages of using a fiber stretcher include lower cost, lower insertion loss and no static frequency shift.
Also, as a group delay actuator, the fiber stretcher is able to directly suppress time of flight fluctuations caused by the atmosphere.
Alternatively, advantages of the AOM include smaller physical size, faster actuation and an infinite phase actuation range (due to acting on the frequency).
However, as a Doppler actuator, the AOM acts on the derivative of the signal phase, and thus the effect on the time of flight will depend in part on the wavelength of different signals.
This has the potential of degrading the efficacy of the phase stabilization system at suppressing time of flight fluctuations.
At the measurement sensitivity of our experiment, this potential degradation was not observed.

In this experiment, the phase stabilization system used a primary optical signal to indiscriminately stabilize independent pure optical signals over a~\SI{7.2}{\tera\hertz} range encompassing the entire C-Band.
However, the indiscriminate phase stabilization may be readily applied to any arbitrary optical signals across the frequency range tested.
Thus, any optical communications signals within C-band may be sent over the link and experience the same level of phase stabilization.
Theoretically, this indiscriminate phase stabilization should also work for optical signals within the S-band, L-band, and beyond.
This suggests that this system could be used to stabilize fiber based communications techniques and equipment for use in free-space communications.
This has the potential to significantly reduce the developmental barriers to entry for designing and utilizing free-space optical communication by leveraging the decades of development in high-speed communications over optical fiber.

Specifically, we note that Kang et al.~\cite{Kang2019} demonstrated that a phase stabilization system, similar to the one presented in this paper, can improve the quality of a~\si{1~Gbps} binary phase-shift keying (BPSK) data stream transmitted over a~\SI{1.4}{\kilo\meter} free-space link.
Therefore, we expect that if a wideband FSO data communication system – such as the~\si{13~Tbps} FSO record, generated using 54 WDM C-band channels, each transmitting~\si{245~Gbps} DP-QPSK~\cite{Dochhan2019} – were enhanced with an ultra-wideband phase stabilization system, the data transmission quality could be improved in cases where atmospheric turbulence limits the transmission.
This will be the case for custom, low SWaP, communication systems that rely on highly-coherent lasers, and where the optical path needs to sweep through an atmospheric slant path to track low Earth orbit satellites.
Based on the current optical fiber-based data rate record demonstrated using a single photonic sources (low SWaP) – 160 WDM wavelength channels, each transmitting at~\si{243~Gbps} 64-QAM~\cite{corcoran2020ultra} – we conclude that~\si{40~Tbps} ground-to-space FSO transmission should be made possible with the combination of our stabilization system and other demonstrated technologies.

\section{Conclusion}
In this letter, we demonstrated a FSO phase stabilization system which uses a primary optical signal (at \SI{193.1}{\tera\hertz} or \SI{1552}{\nano\meter}) to indiscriminately phase stabilize independent optical signals from a tunable laser, spread over a range of~\SI{7.2}{\tera\hertz} encompassing the full optical C-Band.
The phase stabilization servo is closed with both an AOM as a Doppler actuator, and a fiber stretcher as a group-delay actuator.
We show that this indiscriminate phase stabilization is able to suppress phase noise in these independent signals by~\SI{\sim30}{\dB}, down to an average level of~\SI{-39.6}{\dBc\per\hertz}, and~\SI{-39.9}{\dBc\per\hertz} at~\SI{10}{\hertz}, when using the AOM, and stretcher as the actuators.

We demonstrate that the measured suppression is limited by the poor coherence of the secondary optical source, and find the achievable suppression is likely to be more than~\SI{40}{\dB} greater, reaching an average phase noise around~\SI{-90}{\dB\per\hertz} at~\SI{10}{\hertz}. Our analysis indicates that~\si{40~Tbps} transmission should be made possible in the most challenging atmospheric turbulence conditions, with the combination of our stabilization system and other demonstrated technologies.

\section*{Acknowledgments}
This work has been supported by the SmartSat CRC, whose activities are funded by the Australian Government’s CRC Program. The authors wish to acknowledge the helpful contributions from their CRC research partners at The University of South Australia, The Defence Science and Technology Group, Thales Australia, and Thales Alenia Space. Additional thanks go to Maxim Goryachev, Michael Tobar and UWA EQUS node for the use of their tunable laser, and Peter Wolf for his instructive guidance.

\bibliographystyle{IEEEtran}
\bibliography{ref}

% \printbibliography[title={Whole bibliography}]

\end{document}